# Irreversible Heat Flow Across Phase Boundaries in Phase-Separated Manganites

A.L. Lima-Sharma[1], P.A. Sharma[1] and C. Boekema[2]

[1]*Sandia National Laboratories, MS 1314, PO Box 5800 Albuquerque, NM 87123.*

[2] *Dept. of Physics & Astronomy, San Jose State Univ., One Washington Square, San Jose, CA 95192.*

We have investigated the change in entropy with direct measurements of heat flow as a function of magnetic field at fixed temperatures across the entire phase diagram of the phase-separated (PS) compound $La_{0.25}Pr_{0.375}Ca_{0.375}MnO_3$ (LPCMO). At this composition, the compound shows competing charge-ordered/antiferromagnetic (CO/AF) ground states. At a fixed temperature, we observe an increase in hysteresis in the entropy as a function of the applied field. The heat flux shows progressively irreversible hysteresis, which characterizes the energy barriers between the two competing ground states, as the temperature is lowered. The increase in the heat loss correlates with the increase in magnetic viscosity in the phase-separated state.

**Keyword: manganites, avalanche effect, phase transition, heat flow, DSC, entropy.**

Corresponding author:  allimas@sandia.gov

Jo Alloys & Compounds 834 (2020) 155129.  On 10 April 2020, this JAC-Elsevier article was accepted.  Yet, its online publication shows numerous errors:  Cut-off text, missing Figure Captions, incomplete Table ...  None of which is our fault!  This makes our paper hard to read, study and/or understand.  This reproduction is our final (error-free!) revision.  When citing our refereed paper, please also refer to this arXcv print.  Thank you for understanding.

# Introduction

Materials such as the phase-separated (PS) manganites $La_{0.65-y}Pr_yCa_{0.375}MnO_3$ present strong coupling between crystal structure, magnetic ordering and electronic degree of freedom have attracted attention not only for their basic physics properties, but also for their technological applications. In particular, $La_{0.25}Pr_{0.375}Ca_{0.375}MnO_3$ (LPCMO) has a complex temperature applied field phase diagram [1-5]. One unusual feature of this phase diagram is that the conducting ferromagnetic (FM) and insulating charge-ordered (CO) phases coexist over a very wide temperature and magnetic field range. This phase coexistence of FM and CO insulating regions is understood to result from the long-range strain field, reducing the problem to a manifestation of a martensitic structural transformation. At lower temperatures, a type of glass transition mediated by strain has been proposed to occur, which highlights other conditions of equilibrium [2]. This particular composition, with y ~0.275-0.4, has been studied extensively in numerous previous publications, because this is the narrow composition range where ferromagnetic and charge order phase separation has been observed. The phase separation describes in this system seems to have not only phase complexity in terms of the variety of ground states, but also complex dynamics [6-8]. This transition, which is mediated by strain, has been labeled a "strain glass", and has been proposed to occur in a few different materials [8-11]. In this type of glass state, the system is composed of a dynamically phase-separated mixture of two crystal structures. Upon cooling the system does not reach a conventional low entropy state by transforming into one homogenous crystal structure, instead the two different crystal structures coexist, but become frozen and dynamics rapidly cease at some well-defined temperature. While the composition y, within 0.275<y<0.450, changes the transition temperatures for the various phases, the phenomenon of dynamic phase separation does not change qualitatively [3-5].

While the precise nature of this transition is not clear [12], the LPCMO system is an interesting example of this phenomenon, where strain and magnetic order are strongly coupled, so that the magnetic field appears to quench the strain-glass transition. This static and dynamic phase complexity is challenging to understand, but the phase diagram is well studied and highly tunable with applied magnetic field. In other words, it is a convenient model system for studying the strain glass problem.

Measurements on the elastic constants of LPCMO [6] have provided further evidence for a close relationship between strain and the evolution of the PS state. Furthermore, the shear relaxation modulus plays a dominant role in the formation of a network like structure in the intermediate stage of phase separation. While an elastic material strains instantaneously when stretched and returns to their original state once the stress is removed, LPCMO has elements of both of these properties and exhibits time-dependent strain. The phase-separated regions interact through martensitic accommodation, in a state of metastability, which is extremely sensitive to external parameters. The boundaries between phases are not static and the system exhibits dynamic motion, depending on external factors (such as changes in temperature or applied field).

Resonant ultrasound spectroscopy was used to identify phase-separation dynamics at millisecond time scales, which is consistent with later work on transient conductivity measurements on single crystal thin films [13]. Magnetic viscosity measurements reveal a large time-scale dynamic near a specific temperature of 30 K [15]. The reason for this complex behavior in LPCMO is still fascinating and fruitful in the literature [12,13,14-17]. One question is whether the strain-glass transition can be considered a cooperative phase transition, interpreted as an inelastic phenomenon resulting from long-relaxation strain and disorder effects.

We focus on heat flow measurements using differential scanning calorimetry (DSC) across the different phase boundaries and estimate the entropy associated to each phase transition, as well as the irreversibility and reversibility of the transition by quantifying the magnetic field hysteresis. Unlike transport and magnetization measurements, DSC measurements are directly sensitive to changes in the heat flow. The heat flow increases monotonically with decreasing temperature and seems to diverge at the transition that is suspected to undergo a glass transition mediated by strain. This observation, in turn, implies that phase coexistence dynamics are strongly driven by microstructure. Further understanding of the phase coexistence in LPCMO would then require systematic changes in the microstructure, for which there are few studies [17].

**Methods and Materials**

Polycrystalline $La_{0.25}Pr_{0.375}Ca_{0.375}MnO_3$ were synthesized using solid-state reaction methods [1]. Powder samples were sintered in the 1300-1400 °C range for at least 12 hours in order to achieve good crystallinity and ~100-micron grain sizes. This procedure was used to avoid the suppression of phase separation with small particle size [8]. This composition is chosen because it exhibits the largest temperature and magnetic field regime of phase separation. Heat flow measurements were performed in a Quantum Design PPMS platform using two thermoelectric coolers to detect the differential heat flow [18]. The sample was cut and polished in a rectangular shape to ensure homogeneous heat flow through the sample and good thermal contact. A small amount of grease was used to improve thermal contact. The sensitivity, given in Watts per Volts (W/V), of the thermocouple pair was calibrated as a function of the temperature and was found to be insensitive to magnetic fields up to 9 T over the temperature range of 4-300 K. The samples were zero field cooled (ZFC) from room temperature and held to a lower fixed temperature for the measurement. The field was then ramped at a rate of $dH/dt$ = 200 Oe/s for all measurements. Within the limits of ramp rate available for the PPMS, the calculated entropy had no clear dependence on ramp rate of the field. The rate was highly linear in the area of interest. The entropy associated with each transition was calculated by the integral of peak in heat flow as a function of field, eq.1 [11]:

$$\Delta S = \frac{1}{mT} \int_{H_i}^{H_f} \left(\frac{dQ}{dH}\right) dH , \qquad (1)$$

where *m* is the mass of the sample in kg, *T* is the temperature in Kelvin, and *Q* and *H* stand for heat flow and applied magnetic field respectively. $H_i$ and $H_f$ are the initial and final field, respectively. The measured *dQ/dH* data background was subtracted.

## Results and Discussion

Fig.1 shows the heat flow as a function of the applied field at fixed temperatures of 30 and 32 K. At such temperatures, the transition from a static to dynamic phase-separated state is expected to occur according to the phase diagram as the magnetic field is swept. The transition started near 0.7 T for increasing field sweep (closed black circles). The decreasing field sweep (open black circles) presented no remarkable feature indicating that heat was not released and, therefore, the transition is completely irreversible. At T = 32 K, the heat flow as function of field across the boundaries remained irreversible. At T = 76 K, (see Fig.2) the transition from inhomogeneous state to FM state is visible and starts at 1.4 T and ends at 2.5 T (closed black circles). A percentage of energy is recovered in the decreasing field sweep (open black circles). The net entropy extracted from the difference between up and down field curves is 2.85 J/kgK, in a closed magnetic field cycle.

At T = 176 K, the system undergoes in a phase transformation from insulating, charge-ordered state with an antiferromagnetic ground state and an orthorhombic crystal structure into a metallic, ferromagnetic state with cubic crystal structure. This transition starts at ~0.5 T (see Fig.3). The increasing field curve shows a single peak ~ 1 T with entropy of 0.7 J/kgK. For the decreasing field curve, the transition starts and ends approximately at the same field as the increasing field sweep; however, the entropy associated to the transition is higher at 1.36 J/kgK. As the magnetic field is applied, a small amount of energy leads the system into ordering. The transition around 176 K is very broad, and occurs over about 20 K. A broad transition could indicate the slow movement of magnetic boundaries over the time scale of the magnetic field sweep.

At T = 210 K, a transition occurs from a charge-ordered state with no magnetic order into ferromagnetic ordered state. This phase transition is almost reversible, and the entropy loss involved in the process is 0.5 J/kgK. In Table 1, we list the values for fields where a transition starts and ends for each given temperature. The results from Fig.3 and Table 1 indicate that hysteresis observed in the heat flow measurements agree with those measured from magnetization and transport.

Fig. 4 describes the overall behavior of the difference between the entropies |ΔS| obtained increasing magnetic field, $S_{up}$, and decreasing magnetic field $S_{down}$. The difference between the ΔS for the increasing and decreasing field sweeps indicates the irreversible heat loss for the transition at a given temperature. For higher temperatures, the field for which a transition ends in an increasing field curve corresponds quite close to the field where the transition starts at decreasing field, and the maximum value for the heat flow is similar in modulus. Consequently, the |ΔS| is small as compared to |ΔS| at low temperatures. Where there is a large field hysteresis, the

irreversible heat loss (difference between ΔS for up and down sweeps) is large. |ΔS| reveals the irreversibility in the heat flow, upon cycling the magnetic field.

The dynamics inside the phase-separated state in the 76 – 180 K temperature range should be reflected in the entropy changes, if the measurement time scales are slow enough compared to any relaxation behavior. The existence of long-range elastic, magnetic, and charge interactions rise from phase competition and the magnetic field favors ferromagnetic order over charge order. Thermal fluctuations between phases have a time scale proportional to $exp\left(\frac{-E_b}{k_B T}\right)$ where $E_b$ is an activation energy barrier, $k_B$ stands for the Boltzmann constant and $T$ is the temperature. In this scenario, under isothermal conditions, any disturbance in the energy landscape will enable transitions to occur [13]. Quenched disorder in the presence of field will give multiple distinct energy minima, with energy barriers creating hysteresis. An optimal path to realize the phase transition avoids high energy barriers from long-range interactions, keeping the system in the lowest state of energy as possible locally. This condition naturally leads to slow relaxation towards the global free energy minimum. For example, hysteresis behavior and the magnetic field ramping rate dependence are related through avalanche dynamics within the phase-separated state. The driving force for phase separation is proportional to the applied magnetic field since that lowers the free energy of the ferromagnetic phase. A time scale [16] for the field ramp rate can be defined as $\Delta H/\dot{H} = \tau_H$, where $\dot{H} = dH/dt$ is the applied field ramp rate, and $\Delta H$ is defined as the change in the applied magnetic field obtained from the experimental data (see Table 1). $\dot{H}$ is set constant at 200 Oe/s. The non-linear nature of nucleation in the phase-separated state of the system leads to avalanche behavior over power law timescales [13,14], where the transitions are seen as sharp-short events in time followed by long periods of inactivity. The inset of Fig.4 reveals the dependence of $\tau_H$ for applied field at fixed temperature. The magnitude of the $\tau_H$ estimated here comparable to the time scale over which the magnetic relaxation of the phase-separated state occurs [15]. Therefore, these heat flow measurements conditions should be slow enough to be sensitive to phase-separated relaxation effects.

There is a correlation between the reversibility and $\tau_H$. Note that there is no magnetic behavior or phase separation above 100 K and zero fields; therefore, there should be little relaxation or hysteresis, as observed in the heat flow measurements. A nearly completely reversible behavior occurs when there is a field-induced transition from homogeneous CO to FM state. When the system adopts a phase-separated state, irreversibility is seen and the time scale $\tau_H$ becomes longer because the change in field needed to quench the phase-separation is larger.

We now compare the heat flow measurements directly to previously observed magnetic dynamics. Fig.5 compares the magnetization, heat flow, and magnetic viscosity for LPCMO under the same conditions. The rapid increase in heat loss with decreasing temperature correlates well with the increase in magnetic viscosity in the dynamically phase-separated state. The magnetic viscosity is replotted from reference [14] in Fig.5. The upper panel of Fig.5 shows the static magnetization of LPCMO as a function of temperature and indicates when the various phase transitions occur so that they may be compared with the heat flow and the viscosity. In the upper panel, the LPCMO sample was cooled to 4 K in zero magnetic field. A magnetic field of 1 T was then applied, and the sample was measured upon warming. The rapid rise in magnetization occurs near 30 K when the sample enters the dynamic phase separation state. Near 100 K, ferromagnetic domains disappear, and the sample converts entirely to a charge-ordered state. Near 225 K, LPCMO enters a paramagnetic state. The lower panel of Fig.5 compares the heat loss and magnetic viscosity. In the region below T=100 K, an increase in the static magnetization occurs with time at fixed temperature; a logarithmic time dependence was inferred, allowing a viscosity coefficient to be defined [15]. This time dependence suggests that the non-equilibrium phase-separated state relaxes to a ferromagnetic state. The characteristic time scale for relaxation rapidly increases, as quantified by the viscosity coefficient in the lower panel of Fig.5. The temperature dependence of the viscosity in the phase-separated state from 100 K to 30 K is the origin of the term dynamic phase separation [15]. As the magnetic viscosity increases, indicating an increased $\tau_H$, the irreversible heat loss increases. So, the changes in heat loss are correlated to a slowing down of the magnetic relaxation. Therefore, further study of heat-loss measurements should give insight into the nature of the energy barriers that set the time scale of the slow magnetic dynamics.

## Conclusions

In the present work, the heat loss is investigated across the temperature and magnetic field phase diagram for LPCMO. The irreversible heat loss increases in the dynamic phase-separated regime. The rapid increase in heat loss coincides qualitatively with the increase in viscosity. The magnetic viscosity measurements are only valid when ferromagnetic domains are present below about 100 K. So, the temperature dependence of the heat loss, which is sensitive to the transformation to the FM state at higher magnetic fields, diverges from the magnetic viscosity behavior above this temperature. Our primary result is that irreversible heat loss is consistent with the known magnetic phase diagram of this compound and is observed to track the previously observed magnetic dynamics.

Further examination of the heat loss in the phase-separated regime might allow further insight into the nature of the transition from the dynamic phase-separated state into the static phase-separated state. This transition between a dynamic and static phase-separated state has been observed in the LPCMO, Ni-Ti alloys [20], and Fe-Pd alloys [9], systems in analogy with the glass transition in pure spin systems [21] and in solids [22-26]. Strain is suggested to be the mediating variable in realizing this type of glass transition because it is intrinsic to a phase-separated system in the solid state where the two phases have different structures. The question of whether this transition is an example of a glass transition, in that it shares the same physics as spin glasses or the vitreous state is unclear. What is distinctive about these systems is that two phase-separated macroscopic states coexist, and the 'freezing', if it happens, occurs amongst the relative proportions of the volume fraction between the two phases. Thus, there is little configurational entropy change when 'freezing' occurs between the dynamic and static phase-separated systems. Heat capacity methods detect little change in properties across this transition [26]. Measurements where field changes drive heat in and out of the sample can better distinguish between changes in reversible and irreversible heat loss. Understanding the reversibility of phase transitions is also important for tailoring new materials for the development of renewable magneticaloric energy systems with higher efficiency thermodynamic cycles.

## Acknowledgements

Sandia National Laboratories is a multimission laboratory managed and operated by National Technology and Engineering Solutions of Sandia LLC, a wholly owned subsidiary of Honeywell International Inc. for the U.S. Department of Energy's National Nuclear Security Administration under contract DE-NA0003525. The authors would like to acknowledge S.W. Cheong for helpful discussion. Disclaimer: this paper describes objective technical results and analysis. Any subjective views or opinions that might be expressed in the paper do not necessarily represent the views of the U.S. Department of Energy or the United States Government.

# Table 1

**Entropy associated to field changes at fixed temperatures**

|       | Increasing Field | | | Decreasing Field | | |
|-------|--------|--------|-----------------|--------|--------|-----------------|
| T(K)  | $H_s$(T) | $H_f$(T) | Entropy J/kgK | $H_s$(T) | $H_f$(T) | Entropy J/kgK |
| 30    | 0.6    | 2.6    | 12.79           | -      | -      | -               |
| 32    | 0.6    | 4.2    | 9               | 4.2    | 0.6    | -               |
| 76    | 1.9    | 6.7    | 4.3             | 2.1    | 0.11   | 1.45            |
| 170   | 1.59   | 3.10   | 0.56            | 2.5    | 5.2    | 0.43            |
| 176   | 1.38   | 3.2    | 0.7             | 2.0    | 0.2    | 1.36            |
| 180   | 3.0    | 4.7    | 0.34            | 0.2    | 1.5    | 0.33            |
| 210   | 0.8    | 4.9    | 4.9             | 4.9    | 2.2    | 4.5             |

**Figure Captions**

Fig.1 - Heat flow as a function of applied field at fixed temperatures of 30 and 32 K. Increasing field curves are represented by full symbols, while decreasing field data are presented by open symbols.

Fig.2 - Heat flow as a function of applied field at a fixed temperature of 76 K. Increasing field curves are represented by full black circles, while decreasing field data are presented by open black circles.

Fig.3 - Heat flow as a function of applied field at fixed temperatures 170, 176, 180 and 210 K. Increasing field curves are represented by full symbols, while decreasing field data are presented by open symbols.

Fig.4 - |ΔS| as a function of temperature. The inset shows the field ramp-rate time scale, $\tau_H$, as function of the temperature, calculated from the increasing field sweep.

Fig.5 - Upper panel: Static magnetization as a function of temperature. The sample is measured upon warming after zero field cooling. The onset of ferromagnetic (FM) domains is indicated by the arrow. Below about 100 K, FM regions coexist with charge-ordered (CO) regions. Lower panel: The heat loss upon one magnetic field cycle is plotted with the open circles (left scale). The difference in the heat flow observed for the two field sweeps is calculated and labeled as the heat loss. The heat loss increases rapidly below 100 K. The magnetic viscosity (closed symbols, right scale) is replotted from the literature [15] for the same composition and correlates with heat-loss temperature dependence.

Fig.1.

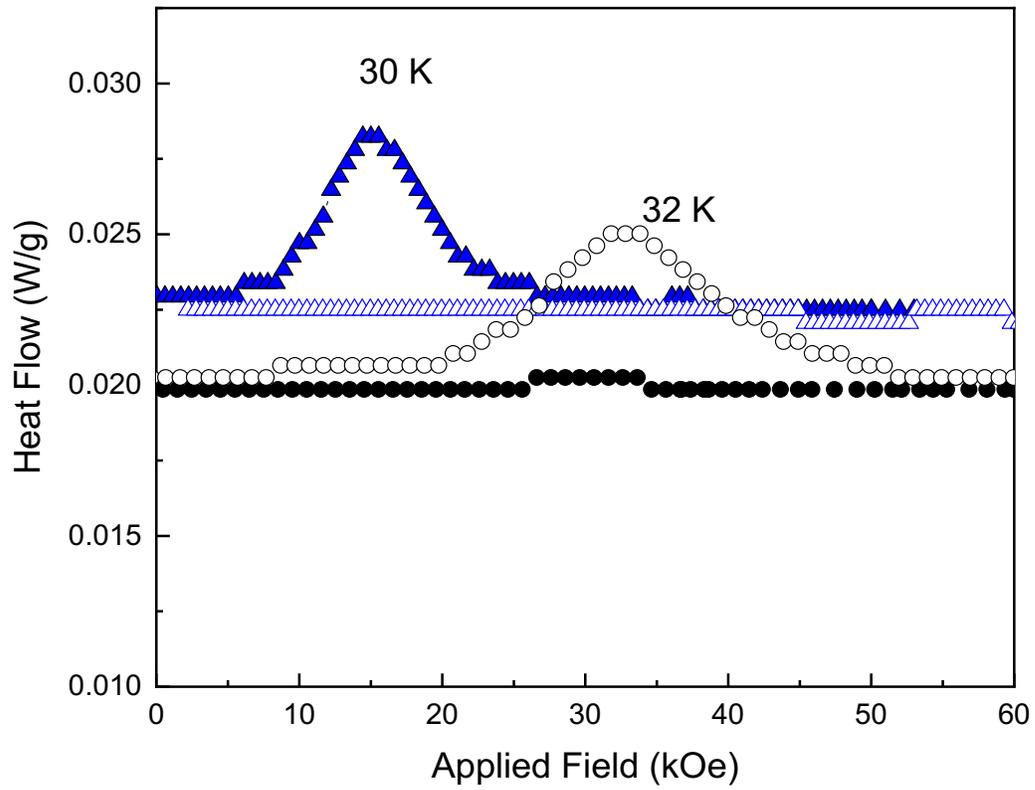

Fig.1 - Heat flow as a function of applied field at fixed temperatures of 30 and 32 K. Increasing field curves are represented by solid symbols, while decreasing field data are presented by open symbols.

Fig. 2.

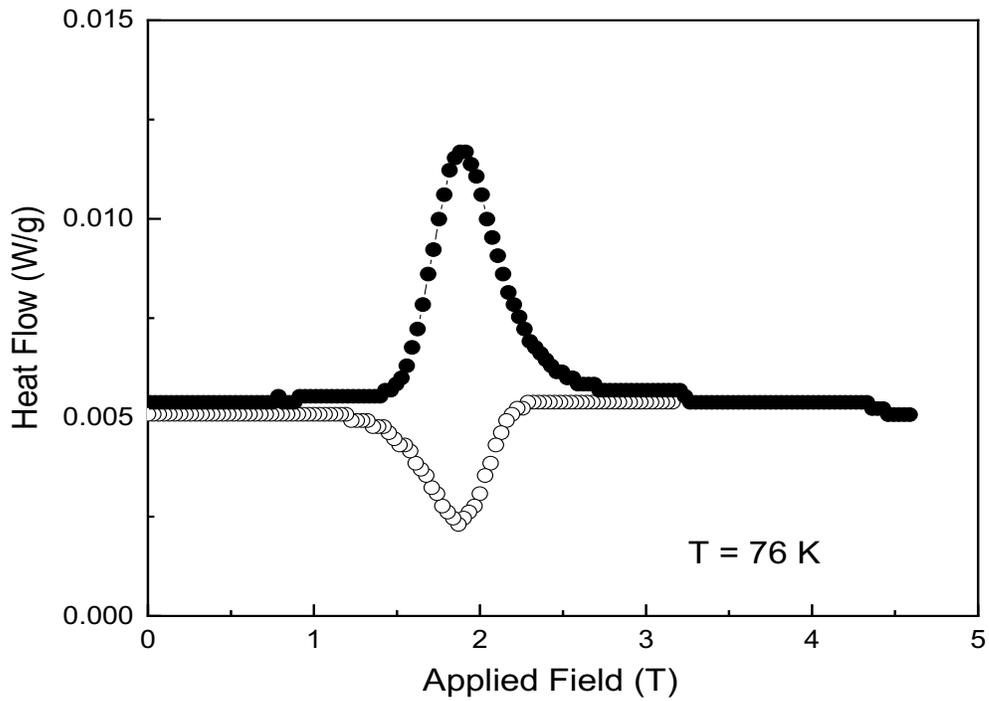

Fig.2 - Heat flow as a function of applied field at a fixed temperature of 76 K. Increasing field curves are represented by solid black circles, while decreasing field data are presented by open black circles.

Fig.3.

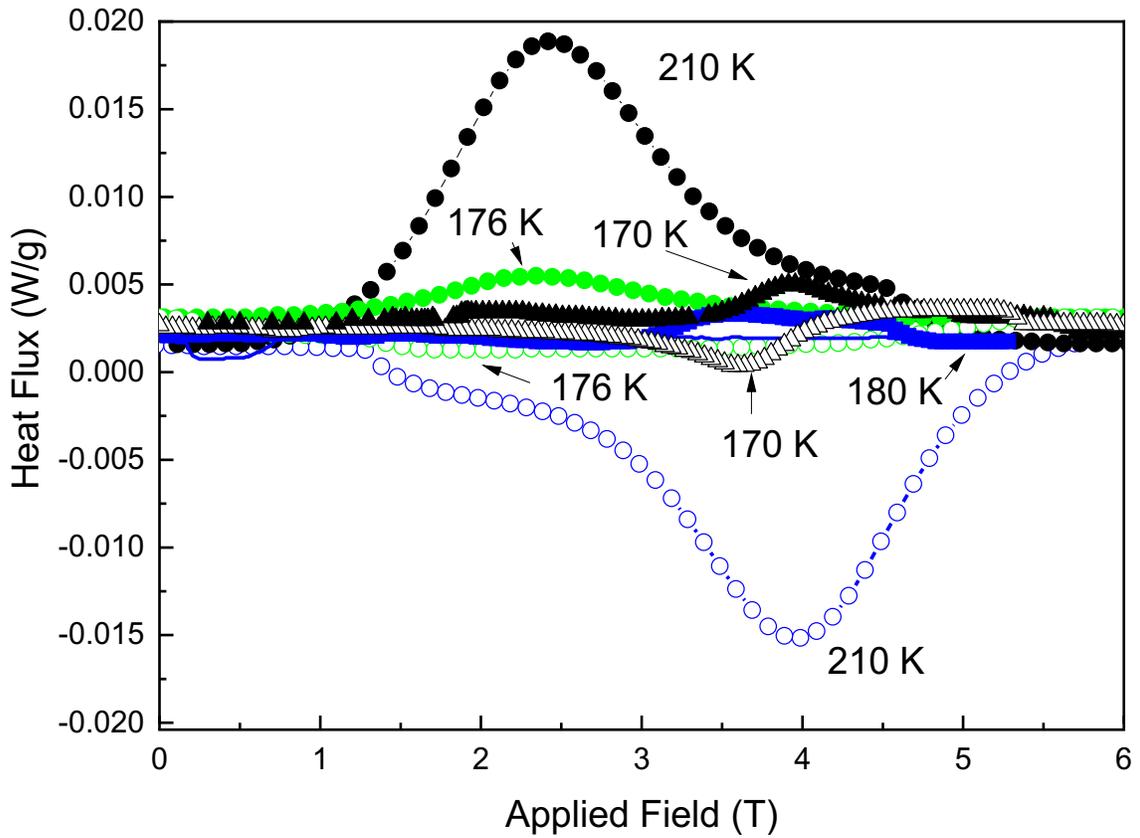

Fig.3 - Heat flow as a function of applied field at fixed temperatures 170, 176, 180 and 210 K. Increasing field curves are represented by solid symbols, while decreasing field data are presented by open symbols.

Fig.4.

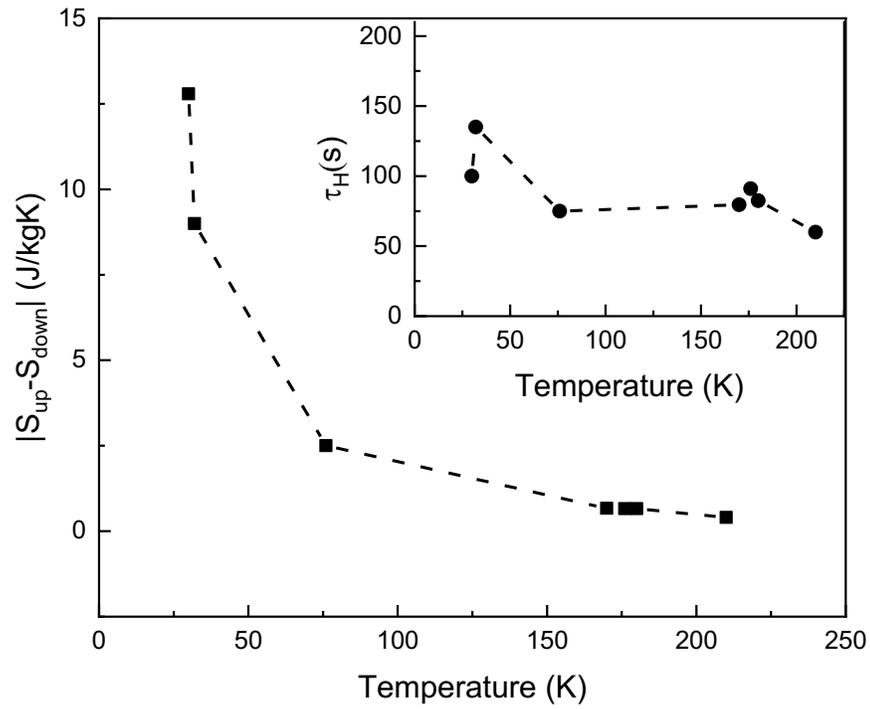

Fig.4 - |ΔS| as a function of temperature. The inset shows the field ramp-rate time scale, $\tau_H$, as function of the temperature, calculated from the increasing field sweep.

Fig.5.

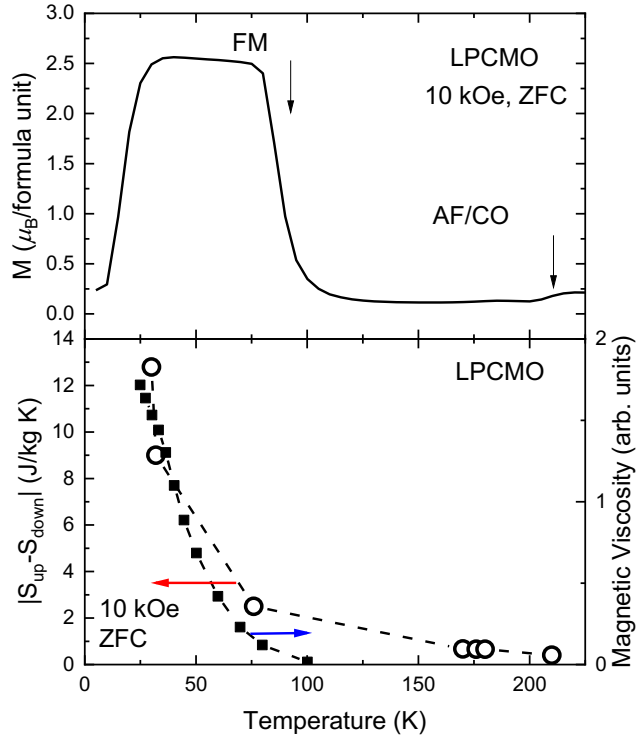

Fig.5 - Upper panel: Static magnetization as a function of temperature. The sample is measured upon warming after zero field cooling. The onset of ferromagnetic (FM) domains is indicated by the arrow. Below about 100 K, FM regions coexist with charge-ordered (CO) regions. Lower panel: The heat loss upon one magnetic field cycle is plotted with the open circles (left scale). The difference in the heat flow observed for the two field sweeps is calculated and labeled as the heat loss. The heat loss increases rapidly below 100 K. The magnetic viscosity (solid symbols, right scale) is replotted from the literature [15] for the same composition and correlates with the heat-loss temperature dependence.